\documentclass[journal]{vgtc}                     % final (journal style)
\usepackage{soul}
\soulregister\cite7 % 针对\cite命令
\soulregister\citep7 % 针对\citep命令
\soulregister\citet7 % 针对\citet命令
\soulregister\ref7 % 针对\ref命令
\soulregister\pageref7 % 针对\pageref命令
\soulregister\cref7 % 针对\pageref命令
\soulregister\Cref7 % 针对\pageref命令

\usepackage{dblfloatfix}
\usepackage{wrapfig, scalerel}

\newcommand*{\icon}[1]{\scalerel*{\includegraphics{#1}}{\strut}}
\newcommand{\tname}{\emph{Quantivine}}
\newcommand{\metaphor}{{\tname} is derived from "quantum circuit" and "vine," with the metaphorical representation of a growing and branching plant symbolizing the system's ability to help users navigate the complex pathways of quantum circuits.}

\newcommand{\added}[1]{{{#1}}}

\usepackage{caption}
\onlineid{1061}

\vgtccategory{Research}

\vgtcpapertype{Applications}

\title{
  Quantivine: A Visualization Approach for Large-scale \\
  Quantum Circuit Representation and Analysis
}

\author{%
  Zhen Wen,
  Yihan Liu,
  Siwei Tan,
  Jieyi Chen,
  Minfeng Zhu,
  Dongming Han,
  Jianwei Yin,
  Mingliang Xu,
  and Wei Chen
}

\authorfooter{
  \item
    Zhen Wen, Yihan Liu, Jieyi Chen, Dongming Han and Wei Chen are with the State Key Lab of CAD\&CG, Zhejiang University.
  	E-mail: \{wenzhen\,$|$\,lyh1024\,$|$\,chenjieyi\_juraws\,$|$\,dongminghan\,$|$\,chenvis\}@zju.edu.cn.
  \item
    Siwei Tan and Jianwei Yin are with the Advanced Computing and System Laboratory, Zhejiang University.
  	E-mail: siweitan@zju.edu.cn, zjuyjw@cs.zju.edu.cn.
    \item
      Minfeng Zhu is with Zhejiang University.
      E-mail: minfeng\_zhu@zju.edu.cn.
  \item Dongming Han is also with Hithink RoyalFlush Information Network Co., Ltd. Zhejiang, China.
  \item Mingliang Xu is with Zhengzhou University.
  	E-mail: iexumingliang@zzu.edu.cn.

  \item Wei Chen is the corresponding author.
}

\abstract{%
  Quantum computing is a rapidly evolving field that enables exponential speed-up over classical algorithms. 
  At the heart of this revolutionary technology are quantum circuits, which serve as vital tools for implementing, analyzing, and optimizing quantum algorithms.
  Recent advancements in quantum computing and the increasing capability of quantum devices have led to the development of more complex quantum circuits.
  However, traditional quantum circuit diagrams suffer from scalability and readability issues, which limit the efficiency of analysis and optimization processes.
  In this research, we propose a novel visualization approach for large-scale quantum circuits by adopting semantic analysis to facilitate the comprehension of quantum circuits.
  We first exploit meta-data and semantic information extracted from the underlying code of quantum circuits to create component segmentations and pattern abstractions, allowing for easier wrangling of massive circuit diagrams. 
  We then develop {\tname}, an interactive system for exploring and understanding quantum circuits. 
  A series of novel circuit visualizations are designed to uncover contextual details such as qubit provenance, parallelism, and entanglement.
  The effectiveness of {\tname} is demonstrated through two usage scenarios of quantum circuits with up to 100 qubits and a formal user evaluation with quantum experts.
  {A free copy of this paper and all supplemental materials are available at \url{https://osf.io/2m9yh/?view_only=0aa1618c97244f5093cd7ce15f1431f9}.}
}


\teaser{
  \centering
  \includegraphics[width=\linewidth]{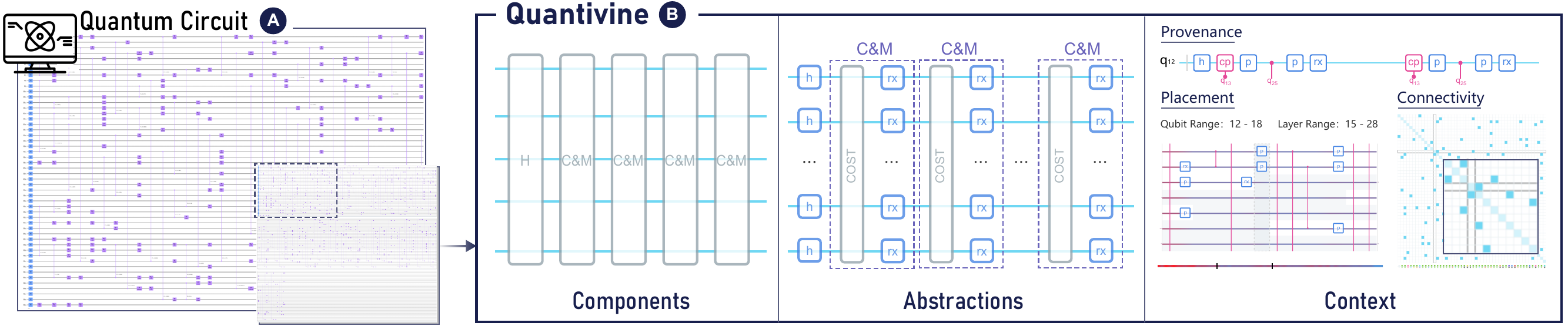}
  \caption{%
    Quantum circuit visualizations for a 50-qubit quantum circuit. 
    (A) Typical quantum circuit diagram generated by Qiskit\cite{Qiskit}. 
    (B) The resulting visualizations of {\tname}, including structured components, pattern abstractions, and context enhancement.
  }
  \label{fig:teaser}
}

\graphicspath{{figs/}{figures/}{pictures/}{images/}{./}{figs/ref/}{figs/icon/}} % where to search for the images

\usepackage{mathptmx}                  % use matching math font

\begin{document}

%%%%%%%%%%%%%%%%%%%%%%%%%%%%%%%%%%%%%%%%%%%%%%%%%%%%%%%%%%%%%%%%
%%%%%%%%%%%%%%%%%%%%%% START OF THE PAPER %%%%%%%%%%%%%%%%%%%%%%
%%%%%%%%%%%%%%%%%%%%%%%%%%%%%%%%%%%%%%%%%%%%%%%%%%%%%%%%%%%%%%%%

\firstsection{Introduction}

\maketitle

\begin{figure*}[tb]
    \centering % avoid the use of \begin{center}...\end{center} and use \centering instead (more compact)
    \includegraphics[width=\linewidth]{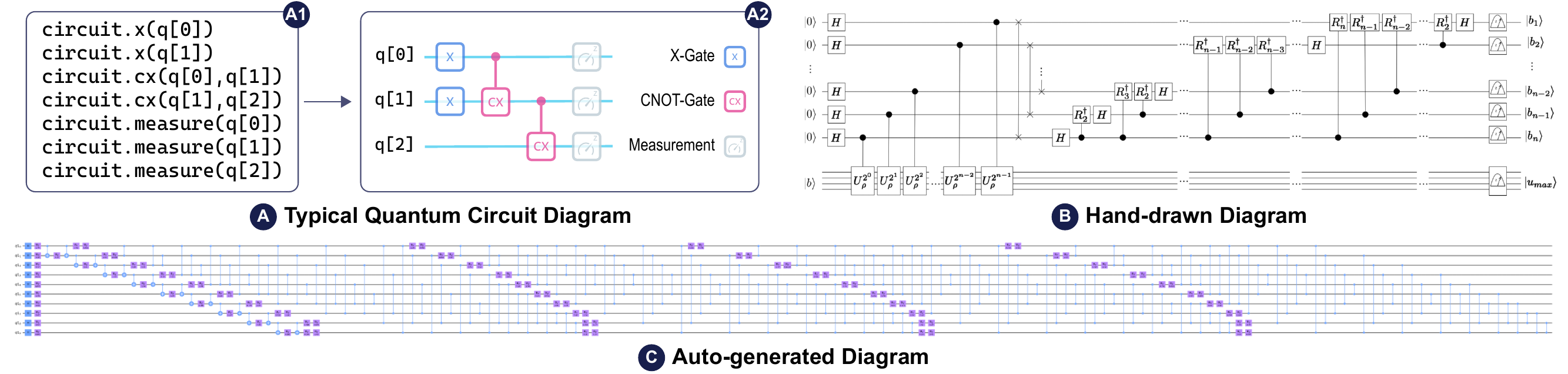}
    \caption{%
        The illustration of quantum circuit diagrams. (A) A typical quantum programming code (A1) and its corresponding quantum circuit diagram {(A2)}.
        (B) A hand-drawn quantum circuit diagram depicts the implementation of quantum PCA algorithm\cite{Martin:2021:TPFD}. (C) An auto-generated diagram (using Qiskit library\cite{Qiskit}) visualizes a quantum circuit having 10 qubits and 306 quantum gates.
    }
    \label{fig:qc-samples}
\end{figure*}

% Quantum computers utilize the principles of quantum mechanics to perform certain computations that are more efficient than classical computers\cite{Nielsen:2010:QCBook}. Most of current available quantum computers use a computation model, called quantum circuit\cite{Deutsch:1989:QCN,Yao:1993:QCC}
% , which is employed for both the representation and implementation of quantum algorithms.
Quantum computers utilize the principles of quantum mechanics, which have the potential to outperform classical computers in certain tasks~\cite{Nielsen:2010:QCBook} when reaching hundreds of qubits.
Most current quantum computers perform calculations by executing quantum circuits, a computation model that is employed for both representation and implementation.
Emergent quantum algorithms implemented by quantum circuits suggest exponential speedup over classical algorithms, which benefits various research communities, such as finance\cite{Egger:2020:QCF}, chemistry\cite{Bauer:2020:QAQCQMS}, biological sciences\cite{Emani:2021:QBS}, and machine learning\cite{Biamonte:2017:QML, Liu:2021:QSML}.

% In the general quantum computing workflow\cite{Weder:2020:IQC}, quantum circuits are defined using quantum programming languages and libraries, such as Q\#\cite{Svore:2018:QSharp} or Qiskit\cite{Qiskit}.
% However, it is well-known that the design of quantum programs is non-intuitive ({\cref{fig:qc-samples}}-A1)\cite{Honarvar:2020:PTQP}.
In the workflow of quantum computing\cite{Weder:2020:IQC}, researchers describe quantum circuits by programming languages or libraries, such as Q\#\cite{Svore:2018:QSharp} and Qiskit\cite{Qiskit} as shown in \cref{fig:qc-samples}-A1. The compiler then transforms the code into executable quantum circuits.
As the original program can not reveal the dependencies between gates,
researchers often use quantum circuit diagrams to assist in the design of quantum algorithms.
These diagrams use time-dependent representations to visualize the structure and behavior of quantum circuits.
% {\Cref{fig:qc-samples}}-A2 shows a typical diagram that depicts a quantum program including the fundamental elements of quantum circuits, \emph{i.e.} \emph{qubits}, \emph{quantum gates}, and \emph{measurements}. 
% Each horizontal line represents a qubit which is the basic unit of information in quantum computing, analogous to a bit in classical computing. 
% The rectangular shapes on the qubits indicate quantum gates which function as operators for qubits, analogous to the role of logic gates in classical digital circuits.
% The sequence of operations follows a left-to-right order.
% Finally, the output of the quantum circuit is observed through the measurement of qubits at the end of the circuit. A well-designed quantum circuit diagram conveys an explicit explanation of a quantum algorithm.
{\cref{fig:qc-samples}}-A2 shows a typical diagram that depicts a quantum circuit including its fundamental elements, \emph{i.e.} \emph{quantum bits} and \emph{quantum gates}.
Each horizontal line in the diagram corresponds to the operation timeline of a quantum bit (or qubit).
A notation (\emph{e.g.} a square box) on the timeline specifies a quantum operation, referred to a quantum gate, that modifies the information stored in the qubits.
The execution sequence of quantum gates follows a left-to-right order.
The final information of qubits after the operations are regarded as the circuit output, which can be sent to other quantum circuits or transformed into classical information by measurement.
In summary, a quantum circuit diagram can explicitly describe a quantum computation procedure.

% However, despite the usefulness of quantum circuit diagrams, 
% Although the usefulness of quantum circuit diagrams has been proven in practice,
% creating clear and effective representations for quantum circuits still poses significant challenges.
% Currently, two techniques are commonly used to visualize quantum circuits, \emph{i.e.}, hand-drawing and automatic generation.
% Hand-drawing methods summarize quantum circuits and present semantic information through visual abstractions (\cref{fig:qc-samples}-B), whereas the manual design and creation process is time-consuming and error-prone.
% % because it requires delicate design and consistency guaranteed between the visualization and the actual quantum circuits.
% It is also tedious to update hand-drawn quantum circuits during the development process to maintain consistency between actual circuits and visualizations. 
% Some quantum computing libraries offer built-in functions for automatic circuit visualization (\cref{fig:qc-samples}-C).
% The automatic generation enables faster and more accurate circuit representation,
% and benefits for the development of small-scale quantum circuits.
The creation of quantum circuit diagrams typically involves two techniques, \emph{i.e.}, hand-drawing and automatic generation.
Hand-drawing methods summarize quantum circuits and present their semantic information through visual abstractions (\cref{fig:qc-samples}-B), whereas the manual design and creation process is time-consuming and error-prone.
Furthermore, updating the diagram when the actual circuit changes is a tedious work.
% It is also tedious to update the diagram when the actual circuit is changed. 
On the other hand, some libraries~\cite{Qiskit, Cirq} offer built-in functions for automatic visualization (\cref{fig:qc-samples}-C), which enable efficient and accurate circuit representation and benefits for the development of small-scale quantum circuits.
However, this technique is limited when scaling to a large number of qubits and quantum gates.
% but can only be scaled to a small number of qubits.
% However, direct printing of whole quantum circuits may result in extremely large outputs, and rendering details illegible.
% These techniques work well for small quantum circuits.

Recent advances in quantum computers pose significant challenges for creating clear and effective representations for quantum circuits,
as the development of quantum circuits are expanding to larger scales.
% As the advances in quantum computers, the design and implementation of quantum circuits are expanding to larger scales.
% % The current state-of-the-art quantum computers have a capability of more than 100 qubits, which enables the implementation of scalable quantum circuits.
% Although the usefulness of quantum circuit diagrams has been proven in practice, creating clear and effective representations for quantum circuits still poses significant challenges.
A number of institutions and universities have proposed quantum computers that can implement circuits with hundreds of qubits.
% , while existing visualization techniques of quantum circuits do not catch up to their pace. 
% To be specific,
% The typical quantum circuit diagrams are inadaptable for rendering large-scale quantum circuits in a limited screen space.
Nevertheless, the increasing number of qubits and quantum gates results in visual clutter and overplotting when they are visualized in typical diagrams.
For instance, {\cref{fig:qc-samples}-C} shows an auto-generated diagram that consists of only 10 qubits and 306 quantum gates, which is hard to examine.
The readability of automatically generated visualizations significantly decreases when scaling up to more than 100 qubits.
% The automatic visualization of a large-scale quantum circuit results in extremely large output, and rendering details illegible.
% The existing techniques have limitations in terms of either scalability or usability.
Therefore, an appropriate visualization approach is required to enable users to easily comprehend and analyze large-scale quantum circuits.

% Recent research in quantum computing applies classical network and graph theory to problems in complex quantum circuits\cite{Biamonte:2019:CNCQ}.
% It presents an opportunity to leverage graph visualization techniques for quantum circuit representation.
This work thus explores how visualization techniques, such as graph summarization\cite{Liu:2018:GSM} and visual abstraction\cite{Viola:2018:PCAV}, can be used for scalable quantum circuits.
Due to the complexity in deciphering quantum circuits, we introduce the semantic information to increase the readability.
To better support the goals of quantum researchers who frequently develop quantum circuits, we conduct expert interviews to understand the pain points and requirements when visualizing quantum circuits.
Inspired by findings from the interviews, we develop {\tname}\footnote[1]{\metaphor}, a {{quant}um c{i}rcuit {vi}sualizatio{n} syst{e}m} that provides efficient visual representations of large-scale quantum circuits and supports interactive exploration and analysis.
% the vast and complex landscape of quantum circuits.
% that generates comprehensible and reasonably detailed quantum circuit visualizations with easy interactions.(\Cref{fig:teaser})

{\tname} utilizes a novel pipeline that extracts latent semantics of quantum programs to automatically visualize scalable quantum circuits using graph summarization and visual abstraction techniques.
Our approach first collects meta-data and semantic information of the quantum circuit from its underlying code.
The circuit is then segmented into components based on its semantic structure, while reorganizing it through qubit bundling and gate grouping.
We also employ a semantic-preserving layout strategy in this process to ensure the new diagram is clear and easy to interpret.
Furthermore, we exploit semantic meanings of the code fragments to identify repetitive patterns in the circuit, which are subsequently presented through visual abstractions.
% These patterns are summarized from a comprehensive survey on current benchmark quantum algorithms.
These patterns are summarized from a comprehensive survey on a circuit benchmark, that covers all well-known quantum algorithms.
Lastly, we design a series of visual representations that complement contextual information to facilitate the comprehension and analysis of the quantum circuit.
% The organization and minimization of quantum circuit representation can be achieved effectively through the use of semantics.

In summary, the contributions of this research include:

\begin{itemize}
    \item We distill a list of design requirements for visualizations of large-scale quantum circuits from expert interviews.
    \item We propose a novel pipeline for visualizing quantum circuits that addresses challenges in readability and scalability with the support of semantic information and a series of visual designs.
    \item We develop {\tname}, a proof-of-concept system to demonstrate the effectiveness of our pipeline through two benchmark quantum algorithms with up to {99} qubits and an evaluation with {10} experts.
\end{itemize}

\section{Related Work}
In this section, we discuss the related work in quantum circuit analysis, semantic analysis, and visualization techniques of complex graphs. 

\subsection{Quantum Circuit Analysis}

% In the current noisy intermediate-scale quantum (NISQ) era\cite{Preskill:2018:NISQ}, 
The quantum circuit is an important computation model of current quantum computers.
% from industry, such as IBM, Google, and D-Wave.
The increasing interest in quantum computing~\cite{Nielsen:2010:QCBook}, coupled with advancements in actual quantum devices\cite{Preskill:2018:NISQ}, has motivated research in the analysis and visualization of quantum circuits.
% Research in analysis, optimization and visualization of quantum circuits is motivated by growing interests in quantum computing and advances in actual quantum devices.

\textbf{Quantum-Classic Difference.}
Although quantum circuits are made up of bits and logical gates similar to classic circuits, they are two different models.
Quantum circuits have unique features such as quantum \emph{superposition} and \emph{entanglement}\cite{Jozsa:2003:REQ, Preskill:2012:QCEF}, which are building blocks of quantum algorithms.
These features power quantum algorithms as necessary parts of any quantum advantage that cannot be replicated in classical circuits.
As such, the research in classical circuits\cite{Kim:2010:EEV} cannot be directly applied to quantum circuits.

\textbf{Performance Analysis.}
Most quantum research leverages mathematical or logical methods during the analysis process of quantum circuits\cite{Maslov:2008:QCS, Bhattacharjee:2017:QCP, Tan:2020:OLS4QC, Weiden:2022:TAS}.
Besides, the network and graph theory is also applied to quantum circuit analysis\cite{Biamonte:2019:CNCQ, Iten:2022:EPPM}.
Recently, VACSEN\cite{Ruan:2023:VACSEN} introduces visualization techniques for quantum circuit analysis,
which provides an intuitive way for users to aware noises in the computation.
These studies mainly focus on the performance of quantum circuits.

% quantum circuits in analysis and optimization process.
% For quantum computing researchers, optimization techniques are usually applied after the design of quantum algorithms.
% Usually, it is imperative to limit the number of gates in the quantum circuit with the purpose of obtaining a robust implementation and meeting the hardware limitations\cite{Maslov:2008:QCS}.
% For example, Tan et al.\cite{Tan:2020:OLS4QC} adapted a more efficient spacetime-based variable encoding, leading to a larger reduction in depth and additional cost.
% \cite{Weiden:2022:TAS}
% On the other hand, an equivalent circuit with lower cost is more favored, 
% \cite{Bhattacharjee:2017:QCP}
% Time-dependent diagrams of quantum circuits describes the process of quantum algorithms.

\textbf{Quantum Circuit Visualization.}
As the growing complexity of quantum circuits and the non-intuitive nature of quantum mechanism, it becomes increasingly challenging to comprehend the circuits.
A number of studies have investigated visualization techniques to aid in the comprehension of quantum computing.
For example, ShorVis\cite{Tao:2017:ShorVis} visualizes a quantum algorithm in terms of quantum states, circuit and probability distribution. 
QuFlow\cite{Lin:2018:QuFlow} displays the parameter flow of quantum circuits.
GraphStateVis\cite{Miller:2021:GSV} offers visual analysis of qubit graph states and their stabilizer.
In addition, Fickler et al.\cite{Fickler:2013:RIQE} present real-time visualization of quantum entanglement state through advanced devices. 
Existing research focuses on visualizing qubit states, parameters or quality indicators to enhance the comprehension of the circuit.
Nevertheless, these studies are limited when scaling to large quantum circuits as the space of quantum states exponential increases.

Our work proposes a novel approach to tackle this challenge by focusing on the efficient representation of quantum circuit structure. We combine semantic analysis and graph visualization techniques to generate visual representations for scalable circuits with high readability. This approach provides a significant contribution towards enhancing the comprehension of quantum circuits and facilitating circuit analysis.
% The application of classical network and graph theory to quantum circuits has shown promising potential\cite{Biamonte:2019:CNCQ}, offering new avenues for the development of efficient and scalable visualization techniques.

\subsection{Semantic Analysis of Programming Languages}
Quantum circuits are commonly constructed using specific programming languages\cite{Svore:2018:QSharp, JavadiAbhari:2014:ScaffCC, Green:2013:Quipper}, or Python toolkits\cite{Qiskit, Cirq}.
% Quantum circuits are commonly constructed using specific programming languages, such as Q\#\cite{Svore:2018:QSharp}, ScaffCC\cite{JavadiAbhari:2014:ScaffCC}, Quipper\cite{Green:2013:Quipper}, or Python toolkits like Qiskit\cite{Qiskit} or Cirq\cite{Cirq}.
Researchers have studied utilizing semantics of quantum circuits for performance optimization and correctness verification\cite{Xu:2022:Quartz, Tao:2022:Giallar}.
But there is a scarcity of research that employs semantics to interpret the quantum circuit.

\added{Semantic analysis has been extensively used in the field of information visualization to facilitate the exploration and analysis of complex datasets.}
Two forms of frequently used semantic information are \emph{semantic structure} and \emph{semantic meanings}.
The semantic structure describes internal relationship of the data, which is often used to navigate exploring data or complex visualizations\cite{Borner:2000:EVSS, Albertoni:2005:VSA, Willett:2007:SW}.
The semantic meanings, such as keywords, are often used to characterize data samples for efficient similarity measurements\cite{Brandes:2002:VBN, Thomas:2004:FPG, Xie:2019:SBM, Wu:2011:SPWC} and visual representations\cite{Sen:2017:Cartograph, Lai:2020:AAS}.
% For example, Lai et al.\cite{Lai:2020:AAS} utilizes semantic analysis on natural languages to annotate text description for visualizations, which enables users to easily understand and interpret complex visualizations.
% \cite{Thomas:2004:FPG, Cao:2010:FacetAtlas}.
In recent years, there has been growing interest in applying semantic analysis techniques to code representation to facilitate comprehension and analysis of programs.
Using semantic structure information, such as abstract syntax tree (AST), to enhance code representation has been proven to be effective\cite{Guo:2021:GraphCodeBERT, Guo:2022:UniXcoder}.
% Moreover, SOMNUS\cite{Xiong:2022:VSD} visualize semantic meanings of data transformation scripts to enhance comprehension of data provenance.
Additionally, certain studies have focused on visualizing the semantic meanings of scripts, aiming to improve comprehension of data provenance\cite{Xiong:2022:VSD, Feng:2023:XNLI}.

\added{Motivated by previous studies, we attempt to utilize semantics to enhance the representation of quantum circuits.
We employ AST to extract the semantic structure of quantum circuits from quantum programs,
followed by inferring semantic meanings of code statements that represents various patterns in the circuits.
Thus, our approach results in highly readable circuit diagrams that incorporate semantics.}

\subsection{Visual Representation of Graph Data}
The visual representation of graph data can be classified into two major categories: node-link diagram and matrix representation.
Matrix representation~\cite{mueller2007comparison,pan2020rcanalyzer,han2022inet} uses an adjacency matrix to visualize a graph and utilizes the ordering of rows and columns to highlight typical patterns.
An intuitive way for non-experts is using node-link diagram by utilizing nodes and edges to represent topology structures~\cite{ghoniem2004comparison,purchase2007important,han2021netv}.
The key issue of node-link diagram is how to place nodes, as the positions impact the visual patterns.
Various layout strategies and methods~\cite{hachul2007large,haleem2019evaluating,kwon2019deep,zhu2020drgraph,Wang:2021:G6} are proposed to handle different tasks and requirements, such as force-directed layout~\cite{hachul2005drawing,jacomy2014forceatlas2}, hierarchical layout~\cite{north2002online}, orthogonal layout~\cite{gorg2005dynamic,biedl1998better} and so on.

As the scale of graph data continues to grow~\cite{lee2006task}, graph summarization\cite{Liu:2018:GSM} and visual abstraction techniques are employed to enhance the readability of graph visualization~\cite{chen2018structure,han2021visual} and facilitate the analysis of graph data. 
Graph summarization focuses on extracting important information from the original graph, 
involving aggregation based, bit compression based, simplification based and influence based methods.
Aggregation based methods aggregate nodes or edges into super-nodes based on optimization function\cite{lefevre2010grass, maccioni2016scalable}.
Bit compression based methods minimize the number of bits in describing graphs\cite{navlakha2008graph, koutra2014vog}.
Simplification based methods remove inessential nodes or edges to simplify graphs\cite{shen2006visual,ghashami2016efficient}.
Influence based methods utilize high-level description of the influence propagation to represent graphs\cite{mehmood2013csi,mathioudakis2011sparsification}.
Some of these techniques may not faithfully represent the original data or introduce complex graph theory concepts that increase the cognitive load on users.
For visual abstraction, it focuses on reducing visual complexity of graphs. Graph motifs\cite{motif2013cody, module2017li} are utilized to simplify graph visualizations, but the motifs and glyphs involve cognitive burden in comprehension.
In addition, customized visual abstraction techniques~\cite{Viola:2018:PCAV} are employed to achieve different tasks and requirements\cite{Zhao:2023:ASTF,Zhou:2019:VALS}.

However, the aforementioned work on visual representation of graph data is extensive and not tailored specifically to quantum circuits.
Given the specialty of quantum circuits, a simple and customized representation that is more suitable for quantum experts is required.
Our work builds upon the requirements of domain experts and provides a significant contribution towards the exploration of utilizing graph visualization techniques for large-scale quantum circuits.

\section{Approach Designs}
In this section, we first introduce the background (\cref{sec:concepts}), then list the design requirements (\cref{sec:requirements}) and our system overview (\cref{sec:overview}).
% Lastly, we present a system overview that demonstrates our pipeline. (\cref{sec:overview}).

\subsection{Background and Concepts}
\label{sec:concepts}
% To aid the understanding of quantum terms and concepts, we have included a glossary in the supplementary materials.
\subsubsection{Quantum Circuit Model}
\label{sec:framework}
Analogous to classical circuits including bits and logic gates, quantum circuits are made up of qubits and quantum gates\cite{Nielsen:2010:QCBook}.
% A general quantum circuit consists of qubits, quantum gates, and measurements.

\textbf{Qubit.}
In quantum computation, quantum qubits, \emph{i.e.} qubits, are basic units of information storage.
A qubit is a ``quantum version'' of a classical bit.
Compared to the classical bit that can be either 0 or 1, a qubit can stay in a \emph{superposition} state\cite{Nielsen:2010:QCBook}.
Mathematically, a qubit state is represented as a linear combination of classical states $|\psi\rangle=\alpha|0\rangle + \beta|1\rangle$, where $|0\rangle$ and $|1\rangle$ represent the classical states 0 and 1, respectively.
Complex parameters $\alpha$ and $\beta$ can be configured via the quantum operation.

% the possible states of a qubit include $|0\rangle$ and $|1\rangle$, like the states 0 and 1 for a classical bit.

\textbf{Quantum gate.}
% In the procedure of quantum computation, the states of qubits are operated by quantum gates.
% The circuit starts with an initial state, which is typically a simple state like $|0\rangle$, and then the gates are applied one after another, transforming the state of the qubits in a specific way.
% For example, a X-gate would inverse the state of a qubit from $|0\rangle$ to $|1\rangle$, like a classical NOT-gate.
% And a CNOT-gate (Controlled-X gate) would inverse the state of a qubit depended on another qubit.
% As shown in {\cref{fig:qc-samples}-A2}, the state of ``q[1]'' is transformed following $|0\rangle \rightarrow |1\rangle \rightarrow |0\rangle$.
% There are various primitive gates used in quantum computing.
% In this research, these gates are unified into two distinct categories: \emph{single-qubit} gates (\emph{e.g.}, X-gate)  and \emph{multi-qubit} gates (\emph{e.g.}, CNOT-gate).
% As the name suggests, a single-qubit gate operates on one qubit at once, whereas a multi-qubit gate interconnects multiple qubits and perform operations on them simultaneously.
% The multi-qubit operations result in the \emph{entanglement} of qubits, which is a crucial factor of quantum advantage\cite{Jozsa:2003:REQ}.
% If two qubits are entangled, their states are logically related and cannot be described independently.
In a quantum circuit, operations on qubits are specified by quantum gates.
Given an initial state of multiple qubits, each gate of the circuit operates one or more qubits, which changes the parameters of qubits~(e,g, $\alpha$ and $\beta$).
For example, a Pauli-X-gate can invert the state of a qubit from $0|0\rangle + 1|1\rangle$ to $1|0\rangle + 0|1\rangle$.
There are many types of quantum gates~(e.g., X-gate and CNOT-gate) introduced in \cite{Nielsen:2010:QCBook}.
In summary, all gate types can be categorized into \textit{single-qubit gates} and \textit{multi-qubit gates}.
A single-qubit gate only operates one qubit, while a multi-qubit gate can operate more qubits, which creates entanglement between qubits~(correlations between parameters of different qubits).
Note that two quantum gates can be executed simultaneously in a circuit if they do not operate the same qubit, leading to parallelism in the qubit timelines.

% \textbf{Measurement.}
% To obtain the output of quantum computation, the quantum circuit is measured, which collapses the superposition of the qubits into a definite state. 
% The measurement result represents the output of computations.
% In general quantum programs, measurements are performed at the end, which do not correlate with the procedure of computations.
% Our research focuses on the representation and analysis of computation procedure that involves only qubits and quantum gates.

% \textbf{Gate composition.}
% Despite of the above primitives, some fixed compositions of quantum gates (\emph{e.g.}, data encoding modules\cite{Schuld:2019:QML, LaRose:2020:RDE}, quantum Fourier transform (QFT)\cite{Ruiz-Perez:2017:QFT}) are reused as components of quantum circuits,
% such as the ``adder'' is reused in the ``multiplier''.
% A functional composition of gates can be seen as a super-gate which has more complex functionality than individual primitive gates.
% In this paper, we refer the composite gates as \emph{component gate}, as opposed to the individual primitive gates.

\subsubsection{Visual Representation of Quantum Circuit}
% \st{To better understand the quantum circuit, a time-dependent diagram is widely adopted by quantum researchers to present its structure and procedure.
% As a common practice in quantum research, a time-dependent diagram is adopted to present the structure and procedure of a quantum circuit for better understanding.
% Our research is based on this general representation.}
% A typical quantum circuit diagram comprises the basic units outlined in the framework (\Cref{sec:framework}).
To interpret the quantum circuit, a time-dependent diagram is widely adopted.
The creation of a quantum circuit diagram involves both glyph representation and layout.
Here we briefly introduce these concepts.
% \hl{Drawing a typical quantum circuit diagram involves designing glyph and layout. Here we briefly introduce its typical visual encoding.}

\textbf{Basic glyphs.}
% \st{The elements of the quantum circuit (outlined in \cref{sec:framework}) are represented as primitive glyphs in the diagram.
% This paper presents them in a uniform visual representation as follows:}
% \hl{Both qubits and gates have visual representation in the circuit diagram:}.
The qubits and quantum gates are represented as glyphs in the diagram.
This paper presents them in a uniform visual representation as follows:
\begin{itemize}
    \item[\icon{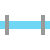}]
     \emph{Qubit wire.}
      A qubit is represented as a horizontal wire in the diagram, %\hl{like storyline?}
      and a set of parallel qubit wires forms the basis of a quantum circuit diagram.

    \item[\icon{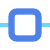}]
     \emph{Single-qubit gate.}
     A single-qubit gate is represented as a square box placed on a qubit wire, which is labeled with the names of gates to differentiate between distinct types of gates, such as ``H'' for a Hadamard gate, ``X'' for a Pauli-X gate.
     % Examples of single-qubit gates include the Hadamard gate and the Pauli-X gate. 
    %   The blocks are labeled with the names of gates to differentiate between distinct types of gates, such as ``H'' for Hadamard gate and ``X'' for Pauli-X gate.
    
    \item[\icon{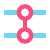}] \emph{Multi-qubit gate.}
      A multi-qubit gate is represented as boxes or dots on the qubit wires, which are vertically connected.
      % Additionally, the boxes and dots indicate different roles of the qubits~(target or controlled). 
      These boxes and dots also serve as an indication of the qubit's role as either a target or controlled qubit. 
      The boxes are also labeled to reflect the specific gate type, such as ``CX'' for a CNOT gate.
      % and horizontally laid on multiple qubit wires, which indicates the gate operates on all connected qubits.
      % An additional control wire with a dot end indicates the control qubit of the gate.
      % Examples of multi-qubit gates include the CNOT gate and the SWAP gate.
    %   The blocks are labeled with gate names as well.
    
    \item[\icon{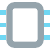}] \emph{Component gate.}
    A component gate is a composition of multiple other gates.
    It is represented as a rectangular box spanning one or more qubit wires.
    The component gate is typically used to simplify the representation of more complex gates in the diagram.
    % For example, the QFT module is typically represented as a component gate composed of several single-qubit and multi-qubit gates in order to simplify its representation.
    % {\color{blue}Figure X} shows an component gate that performs the QFT operation on several qubits.
\end{itemize}
% The boxes of quantum gates are labeled with the names of gates to differentiate between distinct types of gates, such as ``H'' for Hadamard gate, ``X'' for Pauli-X gate, and ``CX'' for CNOT gate.

\textbf{Layout.}
% \st{On the basis of these glyphs, the layout of a quantum circuit diagram conveys specific meanings.
% It provides important visual cues to help readers understand the structure and operation of the circuit.
% For example, the vertical position of a gate on a qubit wire indicates the order in which gates are applied to a qubit, 
% while the horizontal placement of multi-qubit gates indicates the entanglement between qubits.
% In practice, various layout strategies are employed to emphasize specific aspects of the circuit or optimize for certain goals,
% such as minimizing the length of the circuit by positioning all gates to the left of idle wires
% or alleviating parallelism overload by shifting some gates to the right. This work adopts a semantic-preserving layout strategy in order to enhance the readability of the circuit diagram (\cref{sec:segment}).}
The layout provides important visual cues to help understand the circuit.
% For a gate, the notation type corresponds to its operation type.
% Its horizontal and vertical positions specify the operation order and operated qubits, respectively.
Each gate's notation type corresponds to its operation type, while its horizontal and vertical positions specify the order of operations and the qubits being operated upon, respectively.
Notations of multiple-qubit gates connect multiple qubits, indicating potential entanglement between qubits.
The layout can be modified to meet certain goals while keeping the circuit equivalence.
Specifically, the horizontal position of a gate can be moved forward or backward, as long as the order of gates on each qubit remains unchanged.
This work adopts a semantic-preserving layout strategy to enhance the readability of the circuit diagram (described in \cref{sec:segment}).

\subsection{Design Requirements}
% \subsection{Design Requirements}
\label{sec:requirements}
The target users of this work are quantum researchers.
Two domain experts from university labs were involved in the entire process of this research.
In their daily work, quantum circuit diagrams are major visualization tools to interpret quantum circuits.
We conducted iterative interviews with them to distill requirements. 
They propose the requirements for large-scale quantum circuit analysis, which can be translated into visualization requirements below.

\textbf{R1. Clarify the components of quantum circuits.}
When designing quantum algorithms, it is common to use some typical sub-circuits as components of entire circuits.
Quantum researchers could easily recognize a typical component in an isolate circuit.
However, in a complex quantum circuit that consists of numerous quantum gates, it is challenging to identify and distinguish components from the general quantum circuit diagram.
Therefore, new diagrams should clarify the structure of quantum circuits with multiple components.

\textbf{R2. Simplify the patterns of quantum gates.}
Performing batch operations increases the scale of quantum circuits, and results in repeated patterns in circuit diagrams.
In practice, quantum researchers need to identify patterns in diagrams to understand the circuits.
Nevertheless, repeatedly examining patterns imposes a substantial cognitive load.
Therefore, new diagrams should reveal the patterns of quantum gates and reduce unnecessary repetitions.

\textbf{R3. Explicate the context of qubits and quantum gates.}
Quantum researchers have different concerns in quantum circuit analysis, such as qubit provenance\cite{Huang:2019:SAVA}, idling\cite{Das:2021:ADAPT}, quantum gate parallelisms\cite{Figgatt:2019:PEO} and the entanglement of quantum circuits\cite{Hubregtsen:2021:EPQC,JavadiAbhari:2014:ScaffCC}.
\added{These considerations are essential for programming and debugging circuits.
% However, they are implicit in exploring large circuit diagrams because the context of qubits and gates becomes complex and difficult to comprehend.
However, when exploring large circuit diagrams, these contextual details become intricate and challenging to comprehend.
% Therefore, new diagrams should explicate the context information to support idling, parallelism and entanglement analysis of quantum circuits.
Therefore, new diagrams should explicitly present context information, enabling the analysis of idling, parallelism, and entanglement in quantum circuits.}

\textbf{R4. Adopt familiar visual designs and flexible interactions.}
As our users are specialized in quantum computing, they have no experience in visual analysis. 
Thus, a concise and familiar design is preferred.
Also, they need flexibly-customized visualizations on demand.
For example, some users focus on the outline of the circuit, while other users are interested in details of qubits.
New diagrams should thus use visual designs that are intuitive and familiar to quantum researchers.
% , as well as easy to use and customize.

\subsection{System Overview}
\label{sec:overview}
\begin{figure}[htb]
    \centering
    \includegraphics[width=\columnwidth]{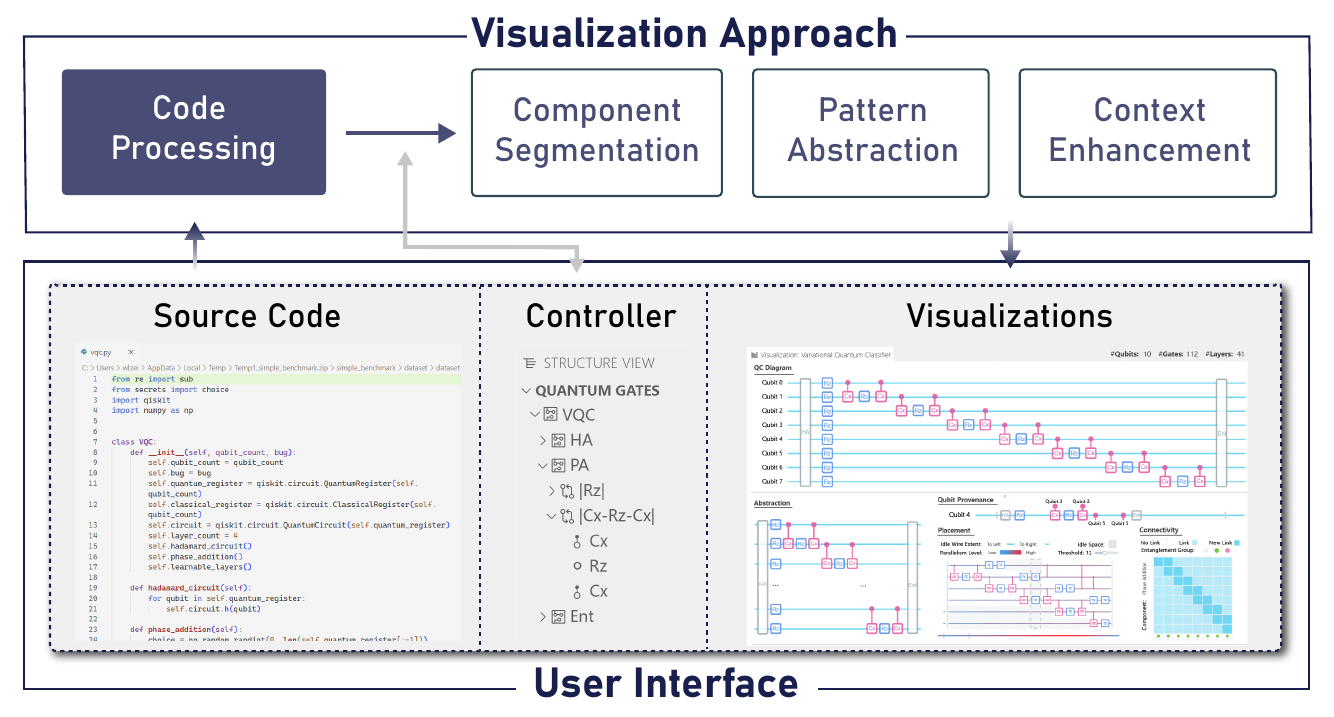}
    \caption{%
    System overview. {\tname} has a four-step visualization pipeline with a three-part interactive interface.%
    }
    \label{fig:system-overview}
\end{figure}

To fulfill these requirements, we design {\tname}, a proof-of-concept system that visualizes scalable quantum circuits.
\Cref{fig:system-overview} illustrates the architecture of the system.
A novel visualization approach (\cref{sec:pipeline}) is undertaken.
It accepts a piece of quantum programming code, and generates flexibly-organized quantum circuit diagrams with comprehensive context information.
A user interface and a series of interactions (\cref{sec:system}) are provided to support interactive exploring and analysis of the quantum circuits.
\added{{\tname} is implemented as a Visual Studio Code plugin and available at \url{https://github.com/MeU1024/qc-vis}.}

\begin{figure*}[tb]
    \centering
    \includegraphics[width=\linewidth]{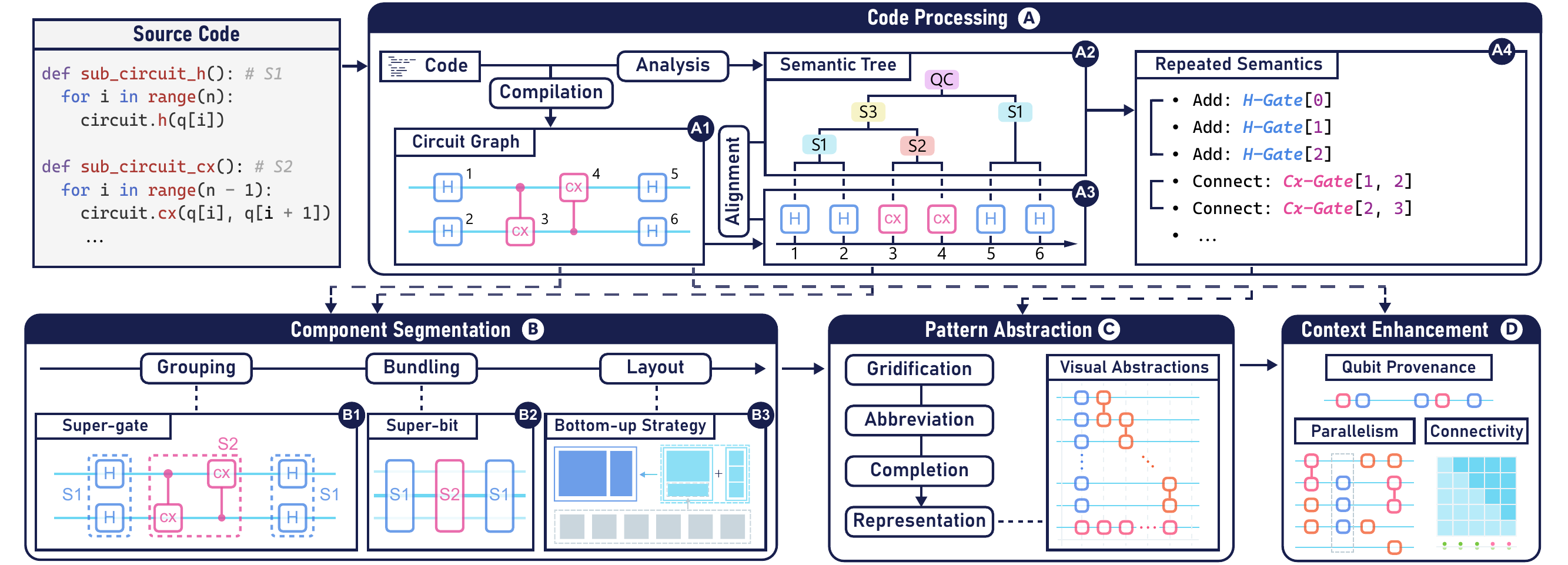}
    \caption{%
        The pipeline of our visualization approach comprises four stages.
        (A) We compile and parse the programming code to collect meta-data and semantic information of the quantum circuit.
        (B) Decompose the circuit to hierarchical components with semantic structure.
        (C) Abstract the circuit using visual abstractions of its patterns.
        (D) Enhance the contextual information with a series of visual representations.
    }
    \label{fig:pipeline-all}
\end{figure*}

\section{Visualization Approach}
\label{sec:pipeline}
In this section, we present a pipeline that accepts the programming code of quantum algorithms, 
and results in a series of visual representations to reveal the components ({R1}), patterns ({R2}) and context information ({R3}) of the quantum circuits (\cref{fig:pipeline-all}).
The semantic structure tree of the quantum circuit is introduced to support flexibly-customized experience throughout the visualization pipeline ({R4}).
To prove the concept of our approach, we implement the proposed pipeline for the quantum circuits that are built using Python with \emph{Qiskit} toolkit. 
% In {\cref{sec:discussion}}, we further discussed that our approach can be easily generalized to other programming languages and toolkits.

\subsection{Code Processing}
\label{sec:code}
% To extract the meta-data and latent semantic information of a quantum circuit, 
% we perform circuit compilation and semantic analysis based on the source code (\cref{fig:pipeline-all}-A).
% Then, we employ data alignment to bridge the gap between the data and semantics.
To support the visualization of a quantum circuit, our pipeline first processes its underlying code through circuit compilation, semantic analysis, and data alignment (\cref{fig:pipeline-all}-A).
This process allows us to extract meta-data and semantic information from the circuit.
% The source code of a quantum circuit may contain condition statements, loops, and function calls, which declares the constructing procedure of the circuit.
% The meta-data pertain to the entities of the circuit, while the semantic information denotes the circuit's structure and patterns present within it.
% \hl{Before visualizing a circuit, we first extract its meta-data from the source code (\cref{fig:pipeline-all}-A), which includes the basic information of the quantum circuit model, like qubits, gates and gate execution orders. To further assist the analysis, we also extract semantic information, which reveals the functionalities of different parts of the circuit.}

\textbf{Circuit Compilation.}
We compile the code and extract the following data from the quantum circuit:
(1) the qubits involved in the circuit,
(2) the quantum gates applied in the circuit with their correlations to qubits, and
(3) the placements of quantum gates in the circuit which indicate the order of execution. 
The above data can be translated to the nodes, edges, and the layout of a circuit graph, which could construct a definite quantum circuit diagram as shown in \cref{fig:pipeline-all}-A1.

% \textbf{Semantic information.}

\textbf{Semantic Analysis.}
Two types of semantic information are derived from the source code.
(1) The \emph{semantic structure} of the code implies the hierarchical structure of sub-circuits.
As there are sub-circuit reuses, researchers usually define some frequently-used circuit construction processes as functions, where each function corresponds to a specific functionality.
Thus, the source code is organized as a tree structure where each tree node represents a function (\cref{fig:pipeline-all}-A2).
We extract this structure through an AST.
% Each function that comprises of a set instructions for building gates corresponds to a sub-circuit, \emph{i.e.} a component gate \icon{component}.
% Thus we get a hierarchical structure of the quantum circuit that consists of a series of component gates.
(2) The \emph{semantic meanings} of code provide insights into patterns in the quantum circuit.
We identify three categories of repetitive patterns from loop statements using a rule-based method, which are detailed in \cref{sec:abs-space}.

\textbf{Node Alignment.}
We apply \emph{node alignment} to establish a connection between the quantum gates and the nodes of semantic structure tree.
Applying semantic or syntactic analysis alone is insufficient to establish a precise correspondence between gates and tree nodes.
We thus insert interrupts in the circuit compilation procedure to track the insertion sequence of quantum gates,
and link the newly-built gates to their semantic representations along with the circuit constructing process. \Cref{fig:pipeline-all}-A3 indicates the \emph{time stamp} of each gate and its corresponding tree node.
As a result, all quantum gates are aligned to the semantic tree nodes.
% For example, ${Gate}_i$ is linked to $Node_i$, which is declared in the function \emph{sub\_circuit\_h}.

% Consequently, the aforementioned data are utilized in the subsequent circuit analysis and visualization process.

\subsection{Component Segmentation}
\label{sec:segment}
To incorporate {R1}, we present a three-step approach for breaking down a typical quantum circuit diagram into hierarchical representations (\cref{fig:pipeline-all}-B).
Our approach draws inspiration from techniques such as node grouping and edge bundling that are effective for graph summarization\cite{Liu:2018:GSM}.
Afterwards, a semantic-preserving layout strategy is introduced to arrange the new diagram.

\textbf{Quantum Gate Grouping.}
The semantic tree (\cref{fig:pipeline-all}-A2) is first employed to guide the segmentation of circuit components.
We label attributes on quantum gates based on a user-customized semantic tree.
Before grouping, users can interactively fold or unfold the tree nodes to control the level of detail (\cref{sec:system:interaction}).
Then, all quantum gates will be labeled with two attributes.
(1) \emph{Tree node label}: 
Each gate will be labeled in accord with the nearest unfolded node to which it belongs.
For example, if the tree node $S1$ is folded while $S3$ is unfolded, then ${Gate}_1$ and ${Gate}_2$ would be labeled with $S3$.
(2) \emph{Loop time label}: 
As a tree node may be repeated in compilation due to loops, two groups of gates may correspond to the same tree node.
We thus label each gate with a time stamp to differentiate separated gate groups that are built from the same function.
Subsequently, the gates will be aggregated by attributes to compose super-gates (\cref{fig:pipeline-all}-B1), analogous to a supernode in a summarized graph.
As a result, the quantum circuit is broken down to a series of primitive \icon{singlegate} \icon{multigate} and component \icon{component} gates.

\textbf{Qubit Bundling.}
% Bundle sequential qubits with the same path hashing.
For a circuit with hundreds of qubits, a typical diagram will display a line for each qubit, which causes massive overlaps when displaying multi-qubit gates.
% A bundled gates involves a number of duplicated wires \icon{qubit} between connected component gates \icon{component}, which may result in visual clutter.
An effective solution to alleviate the problem is adopting the edge bundling technique\cite{Holten:2006:HEB} for combining neighboring edges.
However, the classic method needs to be modified for the quantum circuit.
Due to the fact that each wire represents a distinct qubit throughout the whole circuit,
solely sharing end nodes within a local scope is not a sufficient reason to bundle two qubits together.
They may play a different role in other portions of the circuit.
Therefore, the bundled qubits must be contiguous and of the same provenance throughout the circuit.
The same provenance means these qubits go through the same sequence of primitive or component gates.
For example, all the qubits in \cref{fig:pipeline-all}-B2 go through $S1 \rightarrow S2 \rightarrow S1$, so they can be bundled as one super-bit.

\textbf{Layout Strategy.}
Since the grouping and bundling process breaks up the placements of quantum gates, we present a {bottom-up} layout strategy to reorganize the placements with minimum circuit length and maximum semantic-preserving.
First, we order the sequence of gates by their insertion time stamps (\cref{fig:pipeline-all}-A3). 
The time stamp of the primitive gates \icon{singlegate} \icon{multigate} has been calculated in the \emph{node alignment} process (\cref{sec:code}).
On this basis, the time stamp of the component gate \icon{component} is retrieved from its sub-components.
Secondly, we arrange the layout referring to the semantic structure from the bottom to the top.
The gates in the same tree node are arranged in a local circuit space following the order of their time stamp.
They are laid out on the left most idle place of its correlated qubit wires \icon{qubit} to compress the length of circuit.
If two gates are intersected at the same column, the later gate would be moved backward.
Subsequently, the local layouts of sibling tree nodes will be concatenated while calculating layout for their parent node.
In this way, we could construct a layout from the bottom to the top that shortens circuit length and preserves the semantic structure.

\subsection{Pattern Abstraction}
\label{sec:pattern-abs}
{\tname} abstracts patterns of quantum gates to simplify the representation of the quantum circuit ({R2}).
The pattern of quantum gates refers to the composition and repetition paradigm in a sub-circuit.
However, the composition of the circuit has been explicitly outlined through the component segmentation approach ({\cref{sec:segment}}).
Therefore, we focus on visual abstractions of repeated patterns in this phase.

\subsubsection{Abstraction Space}
\label{sec:abs-space}
\begin{figure}[h]%
    \centering
    \includegraphics[width=\linewidth]{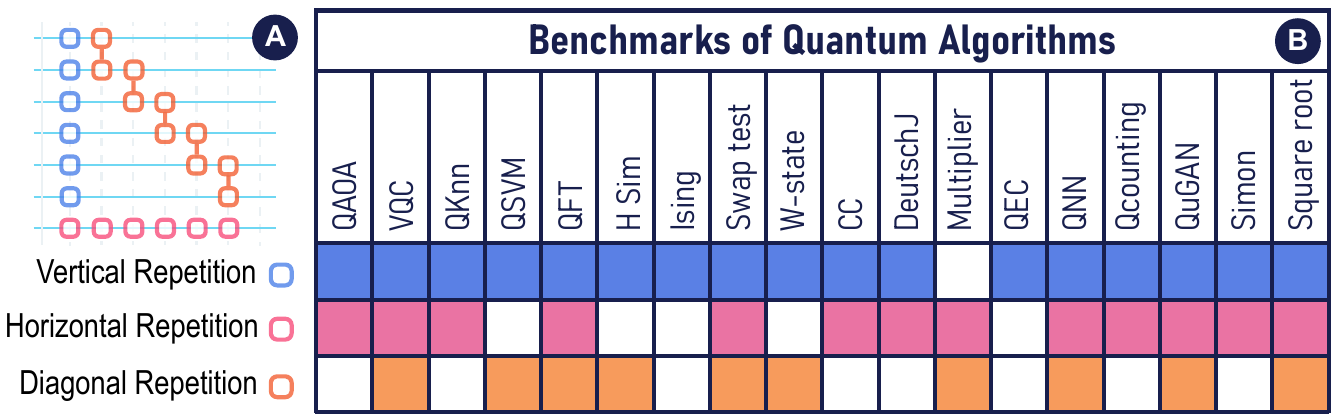}
    \caption{%
        Statistics of the pattern occurrence in 18 benchmark quantum algorithms.
        Three commonly repeated patterns are identified in these algorithms: vertical, horizontal, and diagonal repetition.%
    }
    \label{fig:abs-design}
\end{figure}

Based on a survey of {18} benchmarks of quantum algorithms and expert interviews, we distilled common patterns and abstraction designs of quantum circuits.
The classification and visual representations for these patterns are summarized in \cref{fig:abs-design}.

% \textbf{Directional Repetition Abstraction.}
The statistic result demonstrates that repetitions are necessary patterns in the design of scalable quantum circuits (\cref{fig:abs-design}-A).
% Directional repetitions appear in all 16 benchmarks (15-vertical, 9-horizontal, 9-diagonal).
One observation is that researchers utilize loop statements to repeat sub-circuits for creating scalable circuit.
The repeated sub-circuits are represented as periodic visual patterns in the general quantum circuit diagrams.
We categorize these repetitions owing to their direction, including \emph{vertical}, \emph{horizontal}, and \emph{diagonal} repetitions (\cref{fig:abs-design}-B).

% \begin{itemize}[leftmargin=10pt, noitemsep]
%     \item
\setlength{\intextsep}{0pt}
\setlength{\columnsep}{2pt}
\begin{wrapfigure}{l}{0.5cm}
    \centering 
    \includegraphics[height=0.5cm]{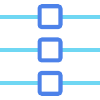} 
\end{wrapfigure}
\noindent
    % When applying the same operation on a sequence of qubits simultaneously, it is shown as a vertical repetition.
    \emph{\underline{Vertical Repetition}} describes the application of the same operation on a sequence of qubits simultaneously.
    A typical instance of vertical repetition is applying H-gates on all qubits simultaneously.

    % \item
    % When applying a sequence of similar operations on a qubit continuously, it is shown as a horizontal repetition.
\setlength{\intextsep}{0pt}
\setlength{\columnsep}{2pt}
\begin{wrapfigure}{l}{0.5cm}
    \centering 
    \includegraphics[height=0.5cm]{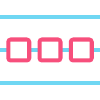} 
\end{wrapfigure}
\noindent
    \emph{\underline{Horizontal Repetition}} presents a sequence of similar operations applied continuously on a single qubit.
    An example of horizontal repetition is applying a series of Rz-gates on the same qubit.

    % \item
\setlength{\intextsep}{0pt}
\setlength{\columnsep}{2pt}
\begin{wrapfigure}{l}{0.5cm}
    \centering 
    \includegraphics[height=0.5cm]{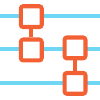} 
\end{wrapfigure}
\noindent
    \emph{\underline{Diagonal Repetition}} stands for applying a sequence of end-to-end operations on a series of qubits continuously.
    These gates are shown in stages.
    An example of diagonal repetition is applying a series of Cx-gates one next to another.

% \end{itemize}
% \st{These repetitive operations can be a sub-circuit which is a composition of quantum gates rather than a single gate.}
In abstractions, each repeated sub-circuit is regarded as a group.
To reveal the frequency and the detail of groups, we preserve the first two sub-circuits and the last sub-circuit for each directional repetition, whilst the intermediate units are simplified as dots. 

\subsubsection{Abstraction Method}
\begin{figure}[h]
    \centering
    \includegraphics[width=\linewidth]{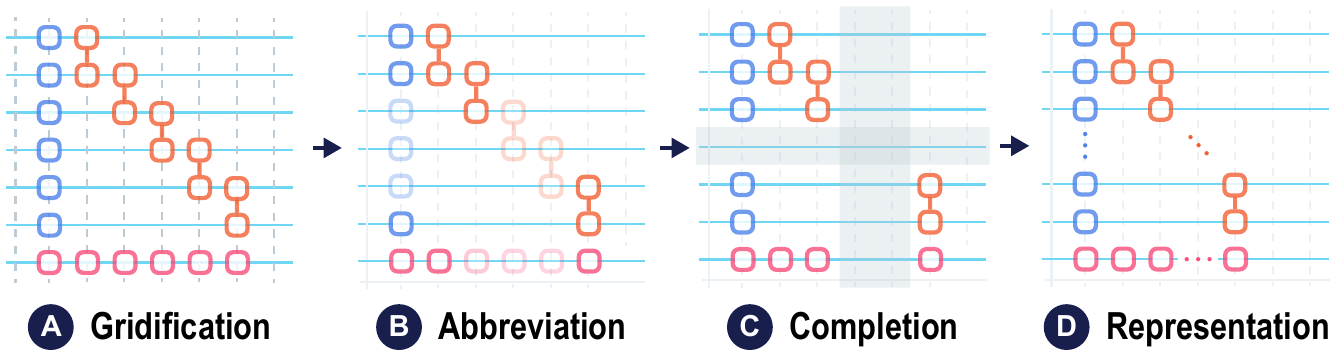}
    \caption{%
        Our abstraction method consists of four steps: (A) gridification, (B) abbreviation, (C) completion, and (D) representation.%
    }
    \label{fig:abs-method}
\end{figure}

Our abstraction method effectively abstracts the quantum circuit by selectively highlighting only the necessary components and removing redundant information.
{\Cref{fig:abs-method} depicts an illustration of the procedure.}
This method consists of the following steps:

\noindent
\textbf{STEP1: Gridification.}
We convert the quantum circuit diagram into a series of grids.
Each grid contains exactly one unit box or none.
We assign a visibility attribute to each row and column of the diagram.
A grid is considered visible if its row or column is visible.
Initially, all grids are invisible. 

\noindent
\textbf{STEP2: Abbreviation.}
We identify repetitive patterns in the quantum circuit with the support of semantic meanings extracted from the code, and highlight their start and end parts.
The covered rows and columns are marked as visible.

\noindent
\textbf{STEP3: Completion.}
We iterate through all quantum gates and determine their visibility.
A quantum gate is visible if and only if its connected qubits are all laid on visible grids.

\noindent
\textbf{STEP4: Representation.}
We render all visible gates and aggregate contiguous idle rows and columns. 
Besides, we complement dot marks in the intermediate space of abstractions to improve readability and produce the final representation of the circuit.

% The circuit visualization is converted into $m \times n$ uniform grids, where $m$ indicates the length of the circuit and $n$ denotes the number of qubits.
% Each grid contains exactly one gate or an idle wire.
% We infer the patterns from the semantic information of the source code.

\subsection{Context Enhancement}
To provide a contextual perspective of the quantum circuit ({R3}), we extract contextual information from meta-data and utilize a set of visual representations to reveal three factors of interest to domain experts.

\begin{figure}[h]
    \centering
    \includegraphics[width=\linewidth]{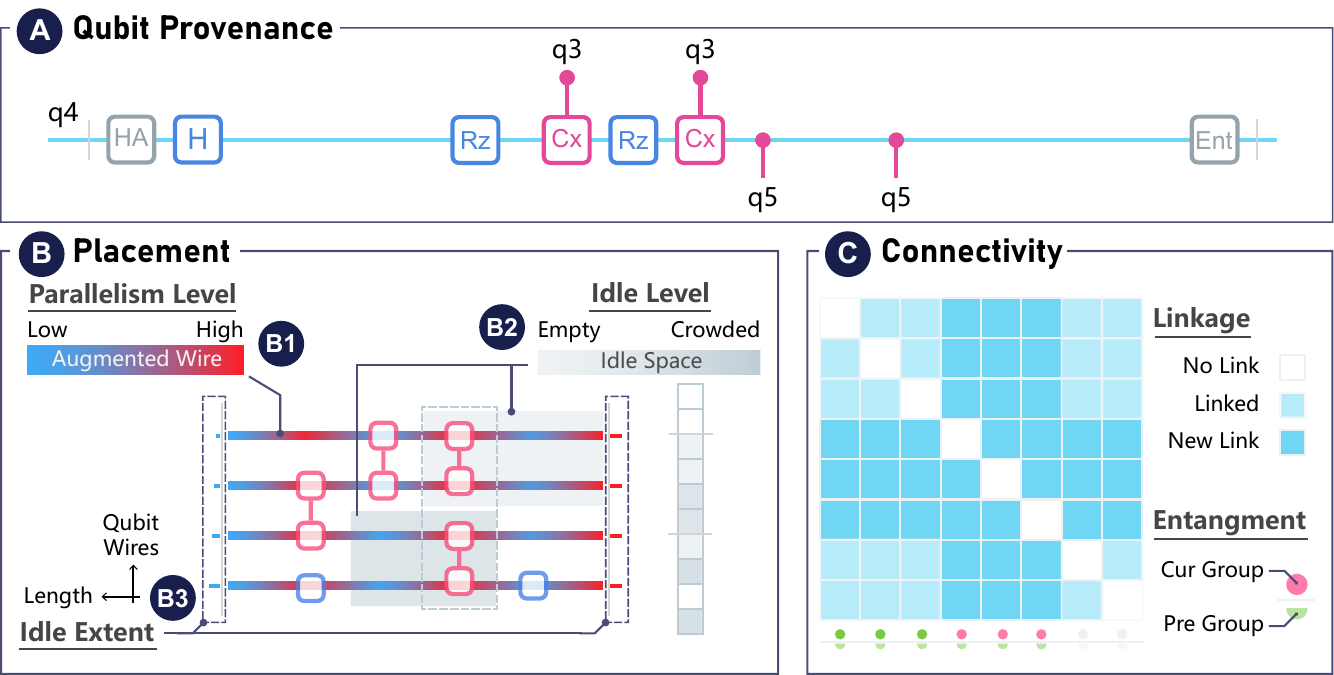}
    \caption{%
        Visual designs of the context visualization.
        (A) The timeline design for qubit provenance.
        (B) The augmented circuit design for quantum gate placement.
        (C) The matrix design for qubit connectivity.
    }
    \label{fig:context-vis}
\end{figure}

\textbf{Qubit Provenance.}
We design a timeline representation for revealing the provenance of each qubit (\cref{fig:context-vis}-A).
All operations applied on the qubit are projected onto the timeline with relative intervals preserved.
% \st{This design facilitates a clear understanding of the genealogy of qubits in the circuit and enables domain experts to trace the origin and evolution of qubits across various stages of the circuit.}
This design enables domain experts to focus on the genealogy of one qubit and trace the evolution across various stages of the circuit.

\textbf{Placement.}
We propose three designs to enhance the visual representation of the quantum circuit and provide contextual information about the placement of quantum gates (\cref{fig:context-vis}-B).
The enhancement involves two aspects: \emph{Parallelism} and \emph{Idle Qubit}.
We first augment the circuit by incorporating color-encoded parallelism levels on the qubit wires (\cref{fig:context-vis}-B1).
Specifically, we assign the color red to indicate heavy-load circuits, while blue denotes light-load circuits.
In addition, the idle spaces around each gate are highlighted to assist in the adjustment of gate placement (\cref{fig:context-vis}-B2). 
The color of the highlighted area represents the idle level.
Notably, when highlighting the idle space for one gate, the idle space of its parallel gates is also highlighted. 
Due to the limitation of screen space, the extent of the idle wire is visualized next to the end of the wires (\cref{fig:context-vis}-B3).
The new diagram reveals the patterns of idling and parallelism in the quantum circuit, which are important factors in understanding and optimizing the performance of the quantum circuit\cite{Bhattacharjee:2017:QCP}.

\textbf{Connectivity.} 
We employ a matrix-based design to depict the qubit connectivity (\cref{fig:context-vis}-C), 
where a highlighted $cell(i, j)$ indicates a direct connection between the $i$-th and $j$-th qubits via one or multiple multi-qubit quantum gates \icon{multigate}. 
Additionally, we propose a glyph-based design to represent entanglement states. The currently entangled qubits are assigned the same color, while the previous entangled states are visually differentiated and reflected below.
This view provides valuable insights into the overall structure and behavior of the circuit, enabling domain experts to make informed decisions about its optimization.

% Overall, the context enhancement process provides a visual and comprehensive perspective on the quantum circuit by revealing important context information of interest to domain experts. 
% This can help researchers and practitioners better understand the behavior and performance of the circuit, and ultimately optimize its design and execution.

\begin{figure*}{
  \centering
  \includegraphics[width=\linewidth]{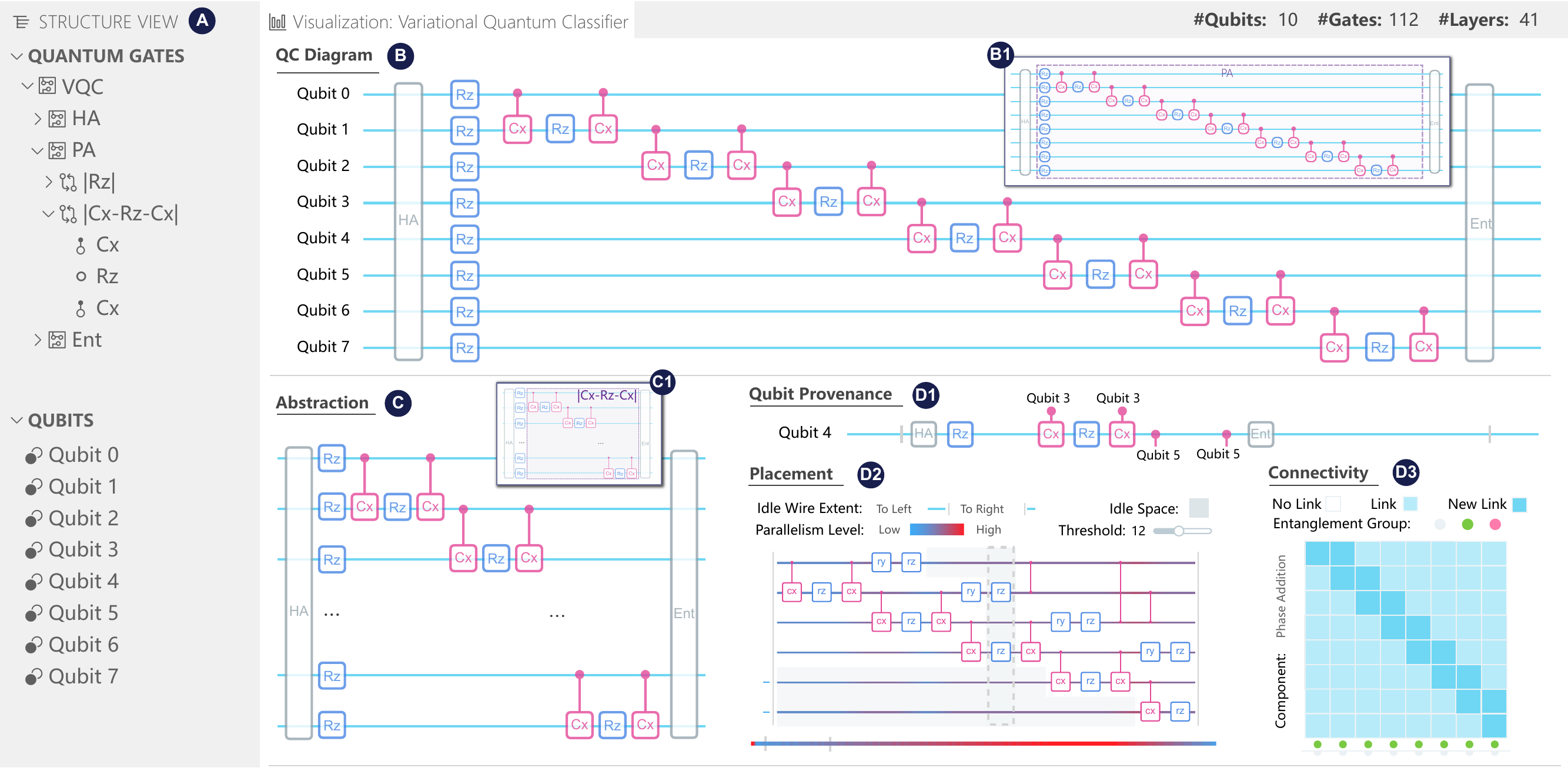}
  \caption{%
    The system interface of {\tname}.
    The Structure View (A) presents the construction of a quantum circuit in a tree diagram including primitive gates, component gates, and repetitive patterns. 
    The Component View (B) provides a flexibly-organized circuit diagram with grouped component gates.
    The Abstraction View (C) shows a further simplified circuit diagram according to visual abstractions of repetitive patterns.
    Three Context Views reveal contextual information in the circuit, including qubit provenance (D1), gate placement (D2), and connectivity (D3).
  }
  \label{fig:system-ui}
}
\end{figure*}

\section{Quantivine: User Interface}
\label{sec:system}
In this section, we introduce the user interface design of {\tname}, which is motivated by our proposed visualization approach,
including the interface design (\cref{sec:system:interface}) and the interaction design (\cref{sec:system:interaction}).
% The system will be used as a technical probe for the usage scenarios (\cref{sec:scenario}) and the user evaluation (\cref{sec:eval}).
\label{sec:ui-design}

% \subsection{System Design}
% The structure view outlines the circuits and acts a part of controller in the system.
% Users use it to navigate in massive quantum circuits, as well as control the organization and the level of details of circuit diagrams.
% Consequently, the visualization view presents the customized visual representations generated by our visualization approach.

\subsection{Interface Design}
\label{sec:system:interface}An overview of the {\tname} interface is shown in \cref{fig:system-ui}, which consists of views (A-D) that present the results of our visualization approach.
{\tname} visualizes the quantum circuit from four aspects:

\begin{itemize}[leftmargin=10pt, noitemsep]
  \item \emph{Structure:}
    The \emph{Structure View} (\cref{fig:system-ui}-A) presents the semantic structure of the quantum circuit as a tree diagram.
    This view serves to provide an overview of the circuit ({R1}) and allows for flexible customization of the circuit visualizations ({R4}). 
    % This view aims to give an overview of the circuit, and serves as a controller to enable flexible customization on the circuit visualizations.
    The tree structure corresponds to the semantic tree extracted from the source code (\cref{sec:code}).
    The branches of the tree outline the hierarchical components and repetitive patterns of the circuit.
    % The components and repetitive patterns of the circuit are hierarchically outlined in the branches of the tree.
    Three interactions based on this view are also designed and further elaborated in \cref{sec:system:interaction}.
    % The components and patterns of the circuit are organized according to collapsible state of each tree item in the structure view.
    % \yezi{User interactions:}
    % We designed three types of interaction, thus enabling users to obtain a better understanding of the circuit and explore it more flexibly (\textbf{R4}).

  \item \emph{Components:}
    The \emph{Component View} (\cref{fig:system-ui}-B) presents a quantum circuit diagram in a generalized form where subsets of quantum gates are aggregated into components.
    This view aims to provide a high-level abstraction of the circuit structure, emphasizing the modularity and reusability of the components ({R1}).
    The components are grouped and arranged based on their execution time and semantics, as extracted from the source code (\cref{sec:segment}).
    The users can use interactions in the \emph{Structure View}, as detailed in \cref{sec:system:interaction}, to customize the level of detail in this view ({R4}).
    % When users completed the design of quantum algorithm, visualizing the circuit which are presented by hierarchical components can provide the pipeline of the algorithm more explicitly. 
    % The gates revealed in structure view will be displayed directly in this part.
    % \yezi{For example, ...}
    % Meanwhile, it is easier for users to spot redundant gate with the assistance of component visualization. \yezi{(fig?)}

  \item \emph{Abstractions:}
    % The \emph{Abstraction View} (\cref{fig:system-ui}-C) presents a more abstract representation of the quantum circuit, where patterns of gates are represented by corresponding visual abstractions.
    The \emph{Abstraction View} (\cref{fig:system-ui}-C) presents the patterns of quantum gates.
    It aims to provide a global perspective of the circuit, highlighting the repetitive patterns and simplifying the visual representation ({R2}).
    % We found that it is highly required to explore the details of quantum circuit which are composed of repetitive patterns most of time. 
    Even though the \emph{Component View} enables scaling down the circuit diagram to a certain extent, it might still take up a large space due to the numerous repeated patterns.
    Hence, we identify and visually abstract such patterns using our approach outlined in \cref{sec:pattern-abs}.
    The level of detail in this view is consistent with the \emph{Component View}.
    % Therefore, abstraction information extracted from the process of pattern abstraction(\Cref{sec:pattern-abs}) will be leveraged to simplify the circuit.\yezi{(fig+example)}

  \item \emph{Context:}
    The \emph{Context Views} (\cref{fig:system-ui}-D1,D2, and D3) consist of three views.
    The first view, D1, provides a summary of the provenance of a qubit within a limited space.
    % First, the provenance of a qubit is summarized in a limited space as D1.
    The second view, D2, depicts the actual placement of quantum gates while visualizing contextual idling and parallelism information.
    % Second, the actual placement of quantum gates are shown in D2, which is visually augmented by contextual idling and parallelism information.
    % Third, the matrix in D3 shows connectivity and entanglement between qubits.
    The third view, D3, displays the matrix showing the connectivity and entanglement between qubits.
    % giving prominence to implicit contextual information in the quantum circuit.
    These views are designed to uncover implicit contextual information within the quantum circuit ({R3}).
    % \st{, to facilitate the detailed analysis of the quantum circuit}
    Additionally, they are interactive and coordinated with other views for specific analysis needs ({R4}).
    % \st{Detailed explanations of the interactive features can be found in \cref{sec:system:interaction}.}
    % We also provided context-oriented visualization. Hence, users can only focus on the gates they are concerned about.\yezi{(fig+example)}
    
\end{itemize}

\subsection{Interactions}
\label{sec:system:interaction}

We implement two interactions in the \emph{Structure View} to enable users to customize the circuit diagrams ({R4}): \emph{folding} and \emph{highlighting}.
% \begin{itemize}[leftmargin=10pt, noitemsep]
\emph{Folding} allows users to fold/unfold items in the \emph{Structure View}.
Initially, all items in the structure tree are collapsed, providing a high-level summary of the circuit. 
Users can expand the items of interest in the tree to explore specific components in more detail. 
This allows users to obtain a more detailed view of the circuit by expanding more low-level items.
With the \emph{highlighting} operation, when users select a particular tree item in the \emph{Structure View}, the corresponding gate will be highlighted in the circuit diagrams in both the \emph{Component} and \emph{Abstraction Views} (\cref{fig:system-ui}-B1,C1). This reduces the manual effort required to match items between code structure and the quantum circuit, and the hierarchical highlights provide a clear division of complex circuit diagrams.
% \end{itemize}

% Users usually find it tedious to map \yezi{code} to its corresponding gate in the large-scale circuit. Highlighting is provided to address this problem.
% \textbf{Focusing.}
% An issue when exploring the details of \yezi{a} certain component in quantum circuit, is that numerous unrelated gates around might distract users. We solved this challenges by providing focusing, which enables users to only pay attention to the components they are interested in since irrelevant part will \yezi{not be displayed(fig.)}.

To facilitate exploration of context information, we design a series of interactions for the \emph{Context Views} in accordance with {R3}.
Users can select a specific qubit in the structure view to display its provenance (\cref{fig:system-ui}-D1).
Clicking on the gates associated with this qubit enables navigation to its placement within the context.
In the placement view, users can adjust the threshold of parallelism level via a scroll bar.
Clicking on a specific column of gates shows potential adjustment places for current parallel operations.
In terms of connectivity, \emph{Quantivine} allows users to specify a component and highlight its connections and entanglement behaviors for a more comprehensive view of the circuit.

\begin{figure*}[tb]
    \centering
    \includegraphics[width=\linewidth]{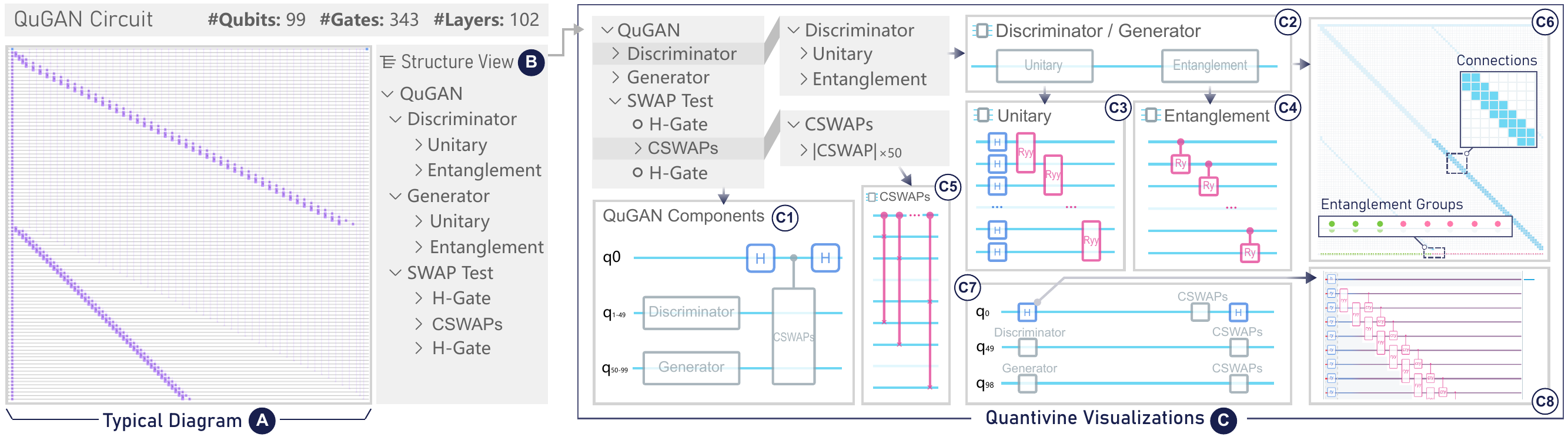}
    \caption{%
        {\tname} generates visualizations based on a 99-qubit QuGAN circuit (A).
        The structure view (B) lists the components of QuGAN.
        User interactions results in visualizations from the component view (C1-2), abstraction view (C3-5) and context views (C6-8).
    }
    \label{fig:case-1}
\end{figure*}

\begin{figure*}[b]
  \centering
  \includegraphics[width=\linewidth]{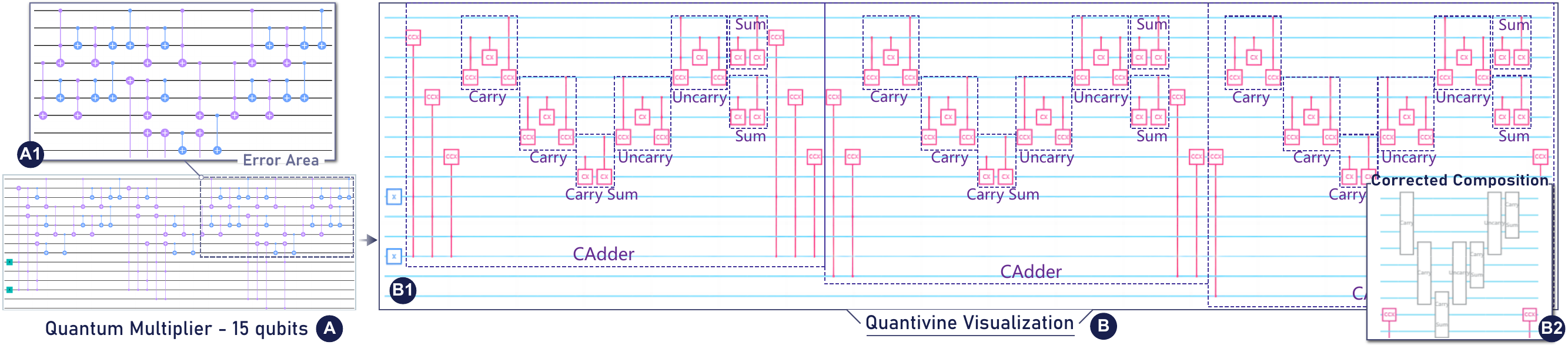}
  \caption{%
    {\tname} wrangles a complex and unstructured quantum circuit diagram (A) into a clear and organized visualization (B) that enables users to easily navigate and identify errors in the circuit.
  }
  \label{fig:case-2}
\end{figure*}

% \subsection{Implementation}
% The system is a webview-based editor plugin written in Typescript, HTML, and CSS using the framework \emph{React JS}.
% To facilitates the efficient rendering, the quantum circuit diagrams are rendered using Canvas, while the other visual cues are rendered as SVGs.
% The input source code is processed using separate Python scripts.
% The visualization results of different quantum programs are included in the supplementary materials.

\section{Usage Scenarios}
\label{sec:scenario}
In this section, we demonstrate the effectiveness of \emph{Quantivine} through two usage scenarios with domain experts.
The first scenario (\cref{sec:scenario-1}) shows the exploration of a 99-qubit quantum circuit that implements a quantum neural network model.
The second scenario (\cref{sec:scenario-2}) introduces the workflow of bug identification in a quantum algorithm.

\subsection{Scenario 1 - Visual Analysis of QuGAN}
\label{sec:scenario-1}
Quantum Generative Adversarial Network (QuGAN) is an emerging topic of quantum machine learning\cite{Stein:2021:QuGAN}. 
However, as shown in \mbox{\cref{fig:case-1}-A}, the meaning of gates in a typical 99-qubit circuit diagram of QuGAN is difficult to understand due to a lack of semantic information.

The expert E1 intends to implement QuGAN on a real quantum computer. Thus, he is interested in exploring the functionality of each part of the QuGAN circuit. 
E1 first examines the high-level structure of QuGAN in the \emph{Structure View} (\cref{fig:case-1}-B): the Discriminator, Generator, and SWAP Test.
After collapsing the components, the circuit diagram becomes clearer and easier to comprehend (\cref{fig:case-1}-C1).
E1 then visualizes the detailed structure of the circuit.
By expanding the Discriminator and Generator components, E1 discovers that they both consist of a Unitary component and an Entanglement component (\cref{fig:case-1}-C2).
Upon further expansion, E1 finds that these components include quantum gates that show repeated patterns.
To confirm the role of these components, E1 employs the connectivity matrix to analyze the connections between different qubits.
As such, E1 leverages the \emph{Abstraction View} to get a concise view of these patterns. In \cref{fig:case-1}-C3, the Unitary component includes a sequence of Ryy gates, connecting qubits in a linear manner. In \cref{fig:case-1}-C4, instead of Ryy gates, the Entanglement component involves linearly-connected CRy gates. E1 says, ``With my knowledge of quantum machine learning, now I understand why the Discriminator and Generator have the same structure. Analogous to classical neural networks, QuGAN trains the parameters of gates to fit the data. In the GAN framework, both the discriminator and generator are neural networks. Thus, their two quantum versions use the same model for simplicity.'' 
To verify this observation, he then highlights the Discriminator, Generator, and SWAP Test components in the matrix view~(\Cref{fig:case-1}-C6), which shows
that the qubits of the Generator component are fully connected. ``This is designed to utilize the unique entanglement property of quantum physics to improve the model complexity, which further improves the prediction accuracy,'' said E1. He also notices that the SWAP Test component entangles all the qubits at the end, as shown in \cref{fig:case-1}-C5, which sends the predictions from the Generator to the Discriminator.

After grasping the detailed structure of the circuit, E1 decides to collapse the details, keeping an overview of it as shown in \cref{fig:case-1}-C1.
He then proceeds to analyze the circuit in terms of the arrangement of qubits and quantum gates.
He first selects q0, and the qubit provenance view shows that it passes through an H-gate, a CSWAPs component, and another H-gate (\cref{fig:case-1}-C7).
This provenance demonstrates that q0 is acting as a control qubit in the SWAP test.
However, via \cref{fig:case-1}-C8, E1 identifies a suboptimal placement of the first H-gate and the CSWAPs component, as there is significant idle time between these operations, which might introduce noise when executed on a quantum computer.
Therefore, he clicks on these two gates to check if it is possible to adjust their placement.
In the placement view, the visual cues show that there is a space for the H-gate to move backwards (\cref{fig:case-1}-C8).
The color of the qubit wire indicates a possible location on the left of the next gate with a low parallelism level, which is suitable for replacing the H-gate. 

% After optimizing the placement of the gates, E1 successfully reduces the total idling time of the circuit.
% With the help of {\tname}, E1 was able to efficiently explore and analyze the large QuGAN circuit diagram, demonstrating the practicality of the system in quantum circuit analysis.

\subsection{Scenario 2 - Bug Detection of Quantum Multiplier}
\label{sec:scenario-2}
The complexity of quantum circuits makes it difficult to detect bugs at the circuit level.
The quantum multiplier, which is designed to calculate the product of two numbers stored in qubits, is a prime example of this complexity.
Even a small-scale quantum multiplier has an intricate structure that is arduous to grasp (\cref{fig:case-2}-A).
% ,making it a challenging task for both novice and expert users to implement a scalable quantum multiplier.

The expert E2 is tasked with detecting bugs in a 15-qubit quantum multiplier circuit, where the bugs are hidden within a cluttered area of the circuit diagram (\cref{fig:case-2}-A1).
To accomplish this, he employs {\tname} to visualize and explore the construction of the circuit.
E2 starts by collapsing all sub-components to obtain an overview of the circuit.
After confirming the high-level structure of the circuit is correct in the \emph{Component View}, he guesses that the bugs are hidden within more detailed structures.
Thus, he drills down by expanding all components.
% Due to benefits of our layout strategy, the fully expanded circuit become though lengthy, it kept an ordered and semantic layout.
Despite the circuit becoming lengthy when fully expanded, our layout strategy ensures an organized and semantically meaningful layout~(\cref{fig:case-2}-B1).
% E2 highlights the components from high-level to low-level and {\tname} provides a more clear and structured representation of the circuit, as shown in \cref{fig:case-2}-B1. 
% He continues to highlight the components from high-level to low-level in order, \emph{Quantivine} provides a clear and structured representation of the circuit, as shwon in \cref{fig:case-2}-B1.
In this view, he notices that the composition of the UnCarry and Sum components in the Adder is incorrect, which should be composed as shown in \cref{fig:case-2}-B2.
Guided by the structured view and visual cues of the circuit diagram, E2 quickly locates the bugs in the underlying code of the circuit and effectively fixes them.

\section{User Evaluation}
\label{sec:eval}
% We conducted a user evaluation to evaluate the effectiveness of our proposed pipeline and visual designs.
In this section, we introduce our evaluation methodology (\cref{sec:eval-method}), and report findings derived from the results (\cref{sec:eval-result}).

\subsection{Methodology}
\label{sec:eval-method}
To evaluate the effectiveness of our pipeline, we conducted a qualitative expert evaluation using {\tname} as a technology probe.
The evaluation had two goals:
firstly, to assess whether our approach is effective in assisting users in comprehending quantum circuits,
and secondly, to evaluate the usefulness of our visual designs in the analysis and optimization of large-scale quantum circuits.

\textbf{Participants.}
We recruited 10 quantum researchers (P1-P10; age: 22-30) from university quantum computing laboratories.
They included undergraduates, graduates, and professors from computer science and physics departments.
All participants majored in quantum computing with experience in developing quantum circuits (P1-P4: < 1 year, P5-P8: 1-4 years, P9-P10: > 8 years).
Their most frequently used visualization tool for quantum circuit is Qiskit.
Four of them (P3-4, P6-7) focused on small-scale circuits with up to 20 qubits, while the others had worked on larger circuits with more than 50 qubits.
We conducted online experiments that lasted 40 to 60 minutes. 
Each participant received approximately \$15 at the start of the session.

\textbf{Task.}
The participants were asked to complete three tasks using {\tname}: a training task (\textbf{T0}), a depiction task (\textbf{T1}), and an analysis task (\textbf{T2}).
For the depiction task, we prepared a quantum circuit with multi-level structure, and required participants to illustrate the backbone of the circuit using the resulting visualizations from the structure, component, and abstraction views.
The analysis task required participants to use context visualizations to analyze and find potential optimization in a quantum circuit.
The training task was prepared to cover all the features in \textbf{T1} and \textbf{T2}.
In total, we prepared six quantum circuits that represents different quantum algorithms, respectively.
Each circuit contains 30 to 99 qubits and over 10 hierarchical components.
We also provided the participants with a document that included the textual description and source code for the algorithms.

\textbf{Procedure.}
The study began with the introduction (10~min) of the study purpose, the motivation of improving quantum circuit representation, and the concepts in our proposed approach.
Next, we proceeded to the training task (\textbf{T0}, 15~min). 
We demonstrated the features of {\tname} with a quantum circuit and asked the experts to reproduce the process themselves.
After the training, we provided the experts with two quantum circuits for the two practical tasks (\textbf{T1} and \textbf{T2}, 10~min for each).
We ensured that the participants were not familiar with the circuit provided in \textbf{T1} and encouraged them to learn and ask questions about the quantum algorithms before starting the trial.
For each task, we asked the experts to depict at least two levels of structure and one visual abstraction.
Finally, the session ended with a semi-structured interview (15~min) and a post-study questionnaire (5-Point Likert Scale). 
% We collected their feedback on our system regarding to advantages, limitations, and potential improvements.
Each session was run in-lab following a think-aloud protocol.

\begin{figure}[tb]
    \centering
    \includegraphics[width=\linewidth]{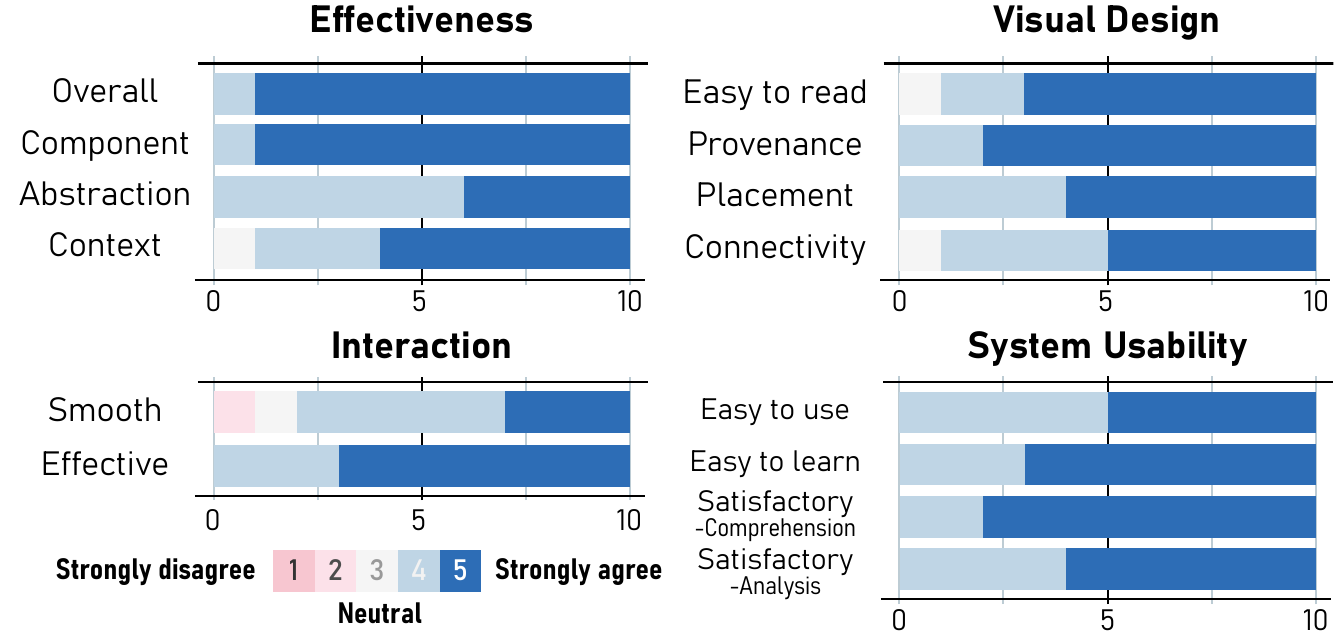}
    \caption{%
        User evaluation results.%
    }
    \label{fig:results}
\end{figure}

\subsection{Findings}
\label{sec:eval-result}
In summary, the evaluation results demonstrate a high level of user satisfaction and effectiveness of {\tname} in supporting quantum circuit exploration and analysis.
\Cref{fig:results} illustrates the results of our evaluation, and we provide a detailed analysis of the findings below. 

\textbf{Effectiveness.}
Participants appreciated the effectiveness of {\tname} in exploring and analyzing quantum circuits ($\mu$ = 4.7, $\sigma$ = 0.5). 
The \emph{Structure} and \emph{Component Views} were particularly praised for their ability to hierarchically outline the circuit, as well as the ability to customize the level of detail in the views.
Participants also noted the usefulness of \emph{Abstraction View} in identifying repetitive patterns, with P8 commenting that \emph{``it's easier to read a scalable circuit with abstractions''}.
Most participants believed the \emph{Context View} provided effective contextual information for circuit analysis, 
whilst some participants expressed the desire for more detailed indicators of the circuit, such as noises and parameters (P1,P3).
% \emph{``want to see more detailed indicators of the circuit, such as noises, parameters, and etc.''} 
Overall, \emph{Quantivine} was perceived as an effective tool for reducing the complexity of circuit representation, and was noted that \emph{``it eases the burden of interpreting circuits''} (P6).

\textbf{Visual Design.}
Participants rated {\tname} highly for its aesthetics, clarity, and usefulness of visual encoding ($\mu$ = 4.6, $\sigma$ = 0.6).
They responded that our design was visually appealing and easy to use.
Participants particularly appreciated the color-coding and highlighting of the placement view for its \emph{``clear and intuitive perception of global placement information''} (P3).
The \emph{Connectivity View} received mixed responses. 
While some participants were \emph{``not convinced''} (P2) due to their unfamiliarity with the matrix representation, other participants found \emph{``it's interesting to investigate the circuit using an adjacency matrix''} (P9) and obtained valuable insights from it.
One participant, P4, specifically noted the usefulness of the connectivity view in his current research about the entanglement analysis of quantum circuit.

\textbf{Interaction \& System Usability.}
Participants responded positively about the usability of {\tname} ($\mu$ = 4.7, $\sigma$ = 0.5), finding it easy to learn and use with fluent and efficient interactions ($\mu$ = 4.4, $\sigma$ = 0.8).
They expressed interest in using {\tname} to comprehend and explain other quantum algorithms in the future, as well as for circuit analysis and optimization.
Furthermore, two professors appreciated for the comprehensive views of quantum circuits provided by \emph{Quantivine}, and expressed interest in sharing the tool in their educational work, indicating its potential to contribute to the advancement of quantum computing education and research.

\textbf{Suggestions.}
Some participants suggested improvements to certain features, such as the ability to annotate and save their work within the tool.
P10 noted that the current system was targeted for the expert users, adding direct manipulation on the diagram, and synchronization with the source code may make it friendly to novice users.
P3 commented that he would be happy if {\tname} could additionally visualize the state of qubits in the intermediate or final stages of the execution.

\section{Discussion}
\label{sec:discussion}
In this section, we reflect on our research and discuss implications for quantum circuit visualization, followed by future work and limitations.

\subsection{Implications}
\label{sec:implication}
% Feedback from our user evaluation with domain experts demonstrates the effectiveness of our pipeline that incorporates semantic analysis and visualization techniques for scalable circuit exploration and analysis.
This study reveals implications for future quantum circuit visualization.

\added{\textbf{Benefits of semantics in quantum circuit interpretation.}}
Our evaluation results indicate that semantics has a promising potential for enhancing the interpretation of quantum circuits.
The user feedback showed a highly positive response towards the semantic representation of the circuit,
which offered a new perspective of circuit visualization.
To our knowledge, this area has not yet been explored in-depth by other researchers. 
This approach effectively reduces the cognitive load of comprehending complex and scalable circuits and further facilitates navigation and analysis tasks.
Moreover, the use of semantics could lead to the development of more advanced tools and techniques for quantum circuit visualization and analysis in the future.

\textbf{Visual representations beyond typical diagrams.}
Our work also explores the potential of visual representations beyond the typical quantum circuit diagrams.
We involve a series of new diagrams to present various perspectives of circuits.
For example, the abstracted circuit diagram allows users to analyze the circuit at various levels of abstraction, making it easier to identify patterns and relationships between different parts of the circuit.
The augmented circuit diagram that uses color-coding and highlighting enhances the perception of gate placement context, which is critical for optimizing the circuit performance.
These new diagrams could be extended to other visualizations and graphical interfaces for quantum computing, providing users with a more intuitive and interactive way to analyze and optimize quantum circuits.

\subsection{Limitations and Future Work}
\label{sec:limitation}
Although our work has shown promising results, there are several limitations and opportunities for future research.

\textbf{Scalability.}
Our semantic-based segmentation and abstraction approaches for quantum circuits are scalable, enabling the visualization of larger and more complex circuits without any limitations.
Our usage scenarios and user evaluation confirm that \emph{Quantivine} can efficiently visualize quantum circuits with up to 100 qubits.
The feedback from domain experts suggests that the system's capability covers the majority of their research on current quantum devices. 
\added{Nevertheless, as circuits scale up to hundreds of qubits, our system encounters challenges with interactions.
The fully expanded circuits and matrix diagrams impose overhead in terms of rendering and visual perception, thereby hindering the interactivity of the system.
Although our design of folding circuit mitigates this issue to some extent, 
optimizing rendering techniques and developing new interactions tailored to the demands of large graphs and matrices can address it more adequately in future work.}

\textbf{Generalizability.}
Although our study focuses on semantic analysis of Python + Qiskit code, the methodologies employed in this research have the potential to be extended to other high-level quantum programming languages.
The fundamental idea involves utilizing static analysis to deduce the structures and meanings of code snippets and then mapping them to dynamic compiled circuits.
These semantic meanings are derived from predefined rules, which can be expanded to encompass more complicated patterns and leverage advanced deep learning techniques for intelligent inference \cite{Wang:2023:VIS} in forthcoming research.

\textbf{Potentiality.}
With increasing capability of the quantum circuit, its visualization and analysis are becoming an emerging field of research.
While our work illustrates a potential pipeline of using semantic analysis and visualization techniques for quantum circuit representation and analysis,
further work is needed to explore how it can be integrated with other quantum programming and simulation tools.
We believe our work has the potential to lay the foundation for future research in large-scale quantum circuits by making their exploration and analysis more accessible to a wider range of researchers and practitioners.

\textbf{Limitations.}
Our system currently supports the visualization of static quantum circuits. 
Future work could focus on developing a real-time visualization system that captures the dynamics of quantum circuits during execution.
The current system is tailored to expert users and lacks certain features that could make it more accessible to novice users.
Future work could focus on improving the user interface and developing features such as direct manipulation on diagrams.
\added{Furthermore, it would be beneficial to offer users the ability to define their own grouping of quantum gates, in addition to automated methods, resulting in a more customizable and user-centric experience.}

\section{Conclusion}
This work presents a novel visualization approach for large-scale quantum circuits, that produces comprehensive representations of the circuit to fulfill the analysis requirements.
The approach involves graph visualization techniques incorporated with semantic analysis, and a set of visual designs specialized for quantum circuits. 
Then, we develop \emph{Quantivine}, a prototype system that allows quantum experts to interactively explore and analyze scalable quantum circuits.
The evaluation results demonstrate the effectiveness of our pipeline and visual designs.
% The findings also suggest that the tool has implications for facilitating the advancement of quantum computing education and research.

%% if specified like this the section will be ommitted in review mode
\acknowledgments{%
This work was supported by the National Natural Science Foundation of China (No. 62132017).
This work was also funded by Zhejiang Pioneer (Jianbing) Project (No. 2023C01036).%
}

\bibliographystyle{abbrv-doi-hyperref}

\bibliography{template}

%% ^^^^^   FOR IEEE VIS, EVERYTHING HERE MAY BE INCLUDED IN THE    ^^^^^ %%
%% 2-PAGE ALLOTMENT FOR REFERENCES, FIGURE CREDITS, AND ACKNOWLEDGEMENTS %%

% \appendix % You can use the `hideappendix` class option to skip everything after \appendix

% \section{About Appendices}
% Refer to \cref{sec:appendices_inst} for instructions regarding appendices.

% \section{Troubleshooting}
% \label{appendix:troubleshooting}

% \subsection{ifpdf error}

% If you receive compilation errors along the lines of \texttt{Package ifpdf Error: Name clash, \textbackslash ifpdf is already defined} then please add a new line \verb|\let\ifpdf\relax| right after the \verb|\documentclass[journal]{vgtc}| call.
% Note that your error is due to packages you use that define \verb|\ifpdf| which is obsolete (the result is that \verb|\ifpdf| is defined twice); these packages should be changed to use \verb|ifpdf| package instead.

% \subsection{\texttt{pdfendlink} error}

% Occasionally (for some \LaTeX\ distributions) this hyper-linked bib\TeX\ style may lead to \textbf{compilation errors} (\texttt{pdfendlink ended up in different nesting level ...}) if a reference entry is broken across two pages (due to a bug in \verb|hyperref|).
% In this case, make sure you have the latest version of the \verb|hyperref| package (i.e.\ update your \LaTeX\ installation/packages) or, alternatively, revert back to \verb|\bibliographystyle{abbrv-doi}| (at the expense of removing hyperlinks from the bibliography) and try \verb|\bibliographystyle{abbrv-doi-hyperref}| again after some more editing.

\end{document}